\documentclass{article}

\usepackage{arxiv}

\usepackage[utf8]{inputenc} 
\usepackage[T1]{fontenc}    
\usepackage{hyperref}       
\usepackage{url}            
\usepackage{booktabs}       
\usepackage{amsfonts}       
\usepackage{nicefrac}       
\usepackage{microtype}      
\usepackage{lipsum}		
\usepackage{graphicx}
\usepackage{doi}
\usepackage{gensymb}
\newcommand\standardstate{{\circ\kern-0.495em-}}
\DeclareUnicodeCharacter{2212}{-}

\title{Unraveling the multivalent Aluminium-ion redox mechanism in 3,4,9,10-Perylenetetracarboxylic dianhydride (PTCDA)}

\author{ \href{https://orcid.org/0000-0002-8490-4239}{\includegraphics[scale=0.06]{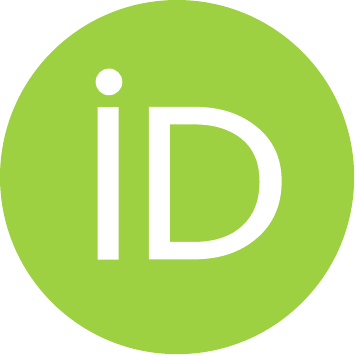}\hspace{1mm}Nicolò~Canever}\\
	School of Information and Physical Sciences\\
	The University of Newcastle\\
	Newcastle, New South Wales, Australia \\
	\texttt{Nicolo.Canever@newcastle.edu.au} \\
	\And
	\href{https://orcid.org/0000-0002-2723-6553}{\includegraphics[scale=0.06]{orcid.pdf}\hspace{1mm}Thomas~Nann} \\
	School of Information and Physical Sciences\\
	The University of Newcastle\\
	Newcastle, New South Wales, Australia \\
	\texttt{Thomas.Nann@newcastle.edu.au} \\
}

\date{}

\renewcommand{\shorttitle}{\textit{arXiv} Template}

\hypersetup{
pdftitle={A template for the arxiv style},
pdfsubject={q-bio.NC, q-bio.QM},
pdfauthor={David S.~Hippocampus, Elias D.~Striatum},
pdfkeywords={First keyword, Second keyword, More},
}

\begin{document}
\maketitle

\begin{abstract}
Rechargeable Aluminium-organic batteries are an exciting emerging energy storage technology owing to their low cost and promising high performance, thanks to the ability to allow multiple-electron redox chemistry and multivalent Al-ion intercalation. In this work, we use a combination of Density Functional Theory (DFT) calculations and experimental methods to examine the mechanism behind the charge-discharge reaction of the organic dye 3,4,9,10-Perylenetetracarboxylic dianhydride (PTCDA) in the 1,3-ethylmethylimidazolium (EMIm$^+$) chloroaluminate electrolyte. We conclude that, contrary to previous reports claiming the intercalation of trivalent Al$^{3+}$, the actual ionic species involved in the redox reaction is the divalent AlCl$^{2+}$. While a less-than-ideal scenario, this mechanism still allows a theoretical transfer of four electrons per formula unit, corresponding to a remarkable specific capacity of 273 mAh g$^{-1}$. However, the poor reversibility of the reaction and low cycle life of the PTCDA-based cathode, due to its solubility in the electrolyte, make it an unlikely candidate for a commercial application. 
\end{abstract}


\section{Introduction}
The global transition to renewable energies is causing a surge in the demand for cheap and dependable energy storage technologies. Although Lithium-ion batteries, the current industry standard for grid-level applications, is a mature technology with reliable performance, there are a number of concerns related to its safety and long-term availability of its rarest elements, such as lithium and cobalt. Therefore, it is essential to look into alternative battery technologies, which are better suitable for applications at the multi-megawatt-hour scale.\cite{armand_building_2008} Among these, non-aqueous Aluminium-ion batteries (AIB) have received some attention by the scientific community in recent years, primarily thanks to the appeal of: (i) low cost and high availability of building block materials, (ii) intrinsic high safety due to the non-flammability of their electrolyte, and (iii) the promise of high energy density thanks to the three-electron redox process of Al metal.\cite{ambroz_trends_2017} This final point, in particular, has been one of the major challenges of the field to date: although the anodic process of aluminium metal electroplating from Lewis-acidic mixtures of Aluminium trichloride (AlCl$_3$) and either ionic liquid-  or eutectic-forming compounds is proven to work reliably,\cite{lai_aluminium_1987,abbott_electrodeposition_2001,jiang_electrodeposition_2006,kitada_alcl3-dissolved_2014,abood_all_2011,abbott_aluminium_2014,canever_acetamide:_2018} a cathode material able to afford high performance and energy density via Al$^{3+}$ intercalation is yet to be reported. While there have been a few sparse publications demonstrating truly trivalent Al$^{3+}$ insertion into inorganic cathode frameworks, \cite{geng_reversible_2015, geng_titanium_2017,hu_binder-free_2018,gu_confirming_2017,canever_cv2o5_2020,he_high-energy_2019}  they are in most cases plagued by poor reversibility, low operating voltages, and lacklustre cycle life/capacity retention. To counteract the unsatisfying results of inorganic cathodes, a new strategy based on organic molecules\cite{kim_rechargeable_2018,bitenc_concept_2020,yoo_tetradiketone_2021,zhou_high-capacity_2021,das_polycyclic_2021} and polymers\cite{raju_self-arranged_2020,wang_high-performance_2020,schoetz_aluminium_2020,vujkovic_polyaniline_2021} has recently been devised. The great advantage of this approach is the virtually unlimited range of compounds that can be designed and used as cathode materials: thus, by using this rationale, a compound allowing the reversible intercalation of trivalent aluminium ions could be ideally created. Among these recent efforts, the case of carbonyl-containing compounds stands out, as it has been demonstrated that these compounds can complex the positively charged AlCl$_2^+$ and AlCl$^{2+}$ ions by forming stable and reversible O-Al-O bonds. \cite{kim_rechargeable_2018,bitenc_concept_2020,yoo_tetradiketone_2021}. This is especially beneficial as it paves the way to: (i) getting closer to a true ''rocking-chair'' intercalation mechanism where Al$^{3+}$ ions are shuttled between anode and cathode without involvement of chloride ions, and (ii) the creation of battery architectures relying on both the intercalation of positive and negative Al-ions, resulting in a better utilisation of the ionic species naturally present in the chloroaluminate electrolyte, and overall higher energy density.\cite{kim_rechargeable_2018,kravchyk_limitations_2020}
Most recently, there have been a few publications reporting the ability of these compounds to afford the intercalation of Al$^{3+}$ ions; \cite{zhou_high-capacity_2021,fang_sulfur-linked_2021} while these results are indeed worthy of excitement, they appear to contradict the works of previous researchers who reported different working mechanism for similar or identical compounds. \cite{kim_rechargeable_2018,bitenc_concept_2020,yoo_tetradiketone_2021} It is therefore evident that a clear understanding of the working principles of such cathode materials has not been achieved yet, and further research is needed to unravel their energy storage mechanism.

In this work, we have targeted such task by examining the redox mechanism of 3,4,9,10-Perylenetetracarboxylic dianhydride (PTCDA) in the 1,3-ethylmethylimidazolium (EMIm$^+$) chloroaluminate electrolyte. PTCDA is an organic dye molecule which has found application in many fields of research, including a variety of emerging energy storage technologies. \cite{luo_organic_2014,chen_organic_2015,fan_ultrafast_2018,han_rise_2020,delaporte_toward_2020,wang_organic_2021} Its molecular structure contains four carbonyl moieties, which are suitable for the complexation with monovalent and multivalent ions; additionally, its n-type semiconductor character makes it an ideal candidate for its employ as a battery electrode. Starting from a theoretical approach using Density Functional Theory (DFT) calculations, and confirming the computed predictions with key electrochemical and material characterisation techniques, we have found that the active aluminium species involved in the discharge-charge process is not the trivalent Al$^{3+}$ as reported in the most recent reports,\cite{zhou_high-capacity_2021,fang_sulfur-linked_2021} but rather the divalent AlCl$^{2+}$, which is aligned with the previous publications.\cite{bitenc_concept_2020,yoo_tetradiketone_2021} Furthermore, we examined the electronic and structural properties of the AlCl$^{2+}$-inserted cathode, and compared it with the case of Al$^{3+}$- and AlCl$_2^+$-inserted structures to better understand its working principle, therefore providing some additional clarity on the topic.

\section{Predicting the redox mechanism of chloroaluminate-inserted PTCDA}
\label{sec:results1}
To predict the correct redox mechanism taking place in the system, we performed a series of DFT calculations using the Quantum ESPRESSO package.\cite{giannozzi_quantum_2009,giannozzi_advanced_2017} Because the active species is a molecular solid, we decided to use a plane wave-based simulation software instead of one employing Gaussian functions for its basis set. By doing this, intermolecular interactions are taken into account, thus giving a more holistic and accurate representation of the system. 
The starting point for the determination of the intercalated structures is an optimised structure of the $\alpha$-polymorph of PTCDA, which was relaxed using the computational settings detailed in the Materials and Methods section, starting from a crystal structure for the molecular solid taken from previous literature.\cite{tojo_refinement_2002} The starting and optimised geometries show overall good agreement (Figure S1), confirming the validity of the computational parameters.

\begin{figure}[h!]
	\centering
	\includegraphics[width=0.99\textwidth]{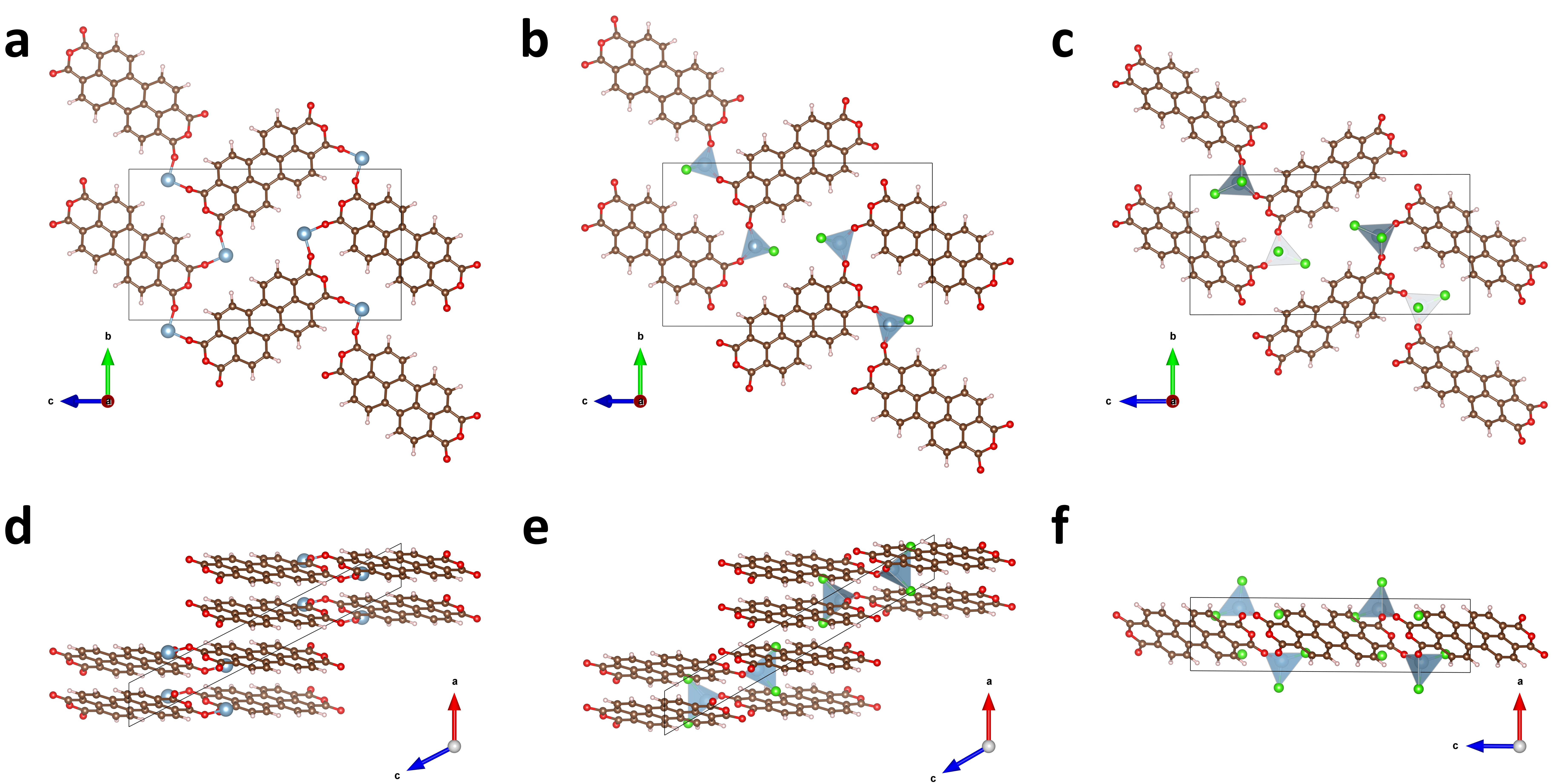}
	\caption{Optimised geometries of PTCDA intercalated with Al$^{3+}$ (a,d), AlCl$^{2+}$ (b,e), AlCl$_2^+$ (c,f), viewed along the \textit{a} (a,b,c) and \textit{b} (d,e,f) axes (colour scheme: C = brown, H = white, O = red, Al = grey, Cl = green). Unit cell sizes are to scale.}
	\label{fig:geometries}
\end{figure}

The Al$^{3+}$- intercalated structure (PTCDA-Al$_2$) was calculated by manually placing Al atoms in key positions of the PTCDA lattice (i.e. symmetrically arranged to allow coordination by the O atoms), and allowing the structure to relax again. All the examined starting points relaxed to the converged structure shown in Figure \ref{fig:geometries} (a,d): the most stable configuration is thus one in which the Al atoms are bridging between two carbonyl O atoms of two adjacent PTCDA units; while the individual units partaking in the Al coordination are initially placed on the same plane, parallel to the (1 0 -3) crystallographic plane, the insertion of Al ions causes the molecules to distort into an angled configuration. Furthermore, the O-Al-O bonds are not co-planar with any of their relative molecular planes, but are also angled with respect to the plane of the perylene moieties. The molecular units, however, still remain stacked in a staggered arrangement along the primary (\textit{a}) crystallographic axis, allowing for the $\pi$-$\pi$-stacking interactions between layered perylene moieties to still take place. This geometry change is also accompanied by a notable distortion and expansion of the unit cell: the \textit{a} and \textit{b} axes are slightly shortened, while the \textit{c} axis is significantly elongated, and the $\beta$ angle is also increased. 

In order to predict the structure of AlCl$^{2+}$- and AlCl$_2^+$- intercalated PTCDA, the starting points for the geometry optimisations were obtained by adding chlorine atoms to the relaxed Al$^{3+}$-inserted structure, co-planar with the O-Al-O bond, at a distance compatible with the formation of a covalent Al-Cl bond. In both cases, the structures relaxed into an configuration in which the Al atoms are at the centre of a tetrahedral coordination environment by two O atoms and two Cl atoms (Figure \ref{fig:geometries} (b,c,e,f)). In the case of the AlCl$^{2+}$- intercalated structure (PTCDA-(AlCl)$_2$), this is made possible thanks to the Cl atoms forming a bridging Al-Cl-Al bond, similarly to the Al$_2$Cl$_7^-$ chloroaluminate ion. the intermolecular distance between PTCDA molecules is reduced, leading to a stronger non-covalent $\pi$-$\pi$-stacking interaction. Conversely, in the case of the AlCl$_2^+$- intercalated structure  (PTCDA-(AlCl$_2$)$_2$), the high steric encumbrance caused by the presence of two distinct Cl atoms in each Al coordination site causes a severe expansion of the unit cell (from 766.45 to 1355.73 {\AA}$^3$), and increase of the interplanar distance of individual molecules. Nevertheless, both AlCl$^{2+}$- and AlCl$_2^+$- intercalated structures present a disruption of the parallel arrangement of PTCDA molecules with the (1 0 -3) crystallographic plane, resulting in a staggered configuration with two sets of parallel molecules.

\begin{figure}[h!]
	\centering
	\includegraphics[width=0.99\textwidth]{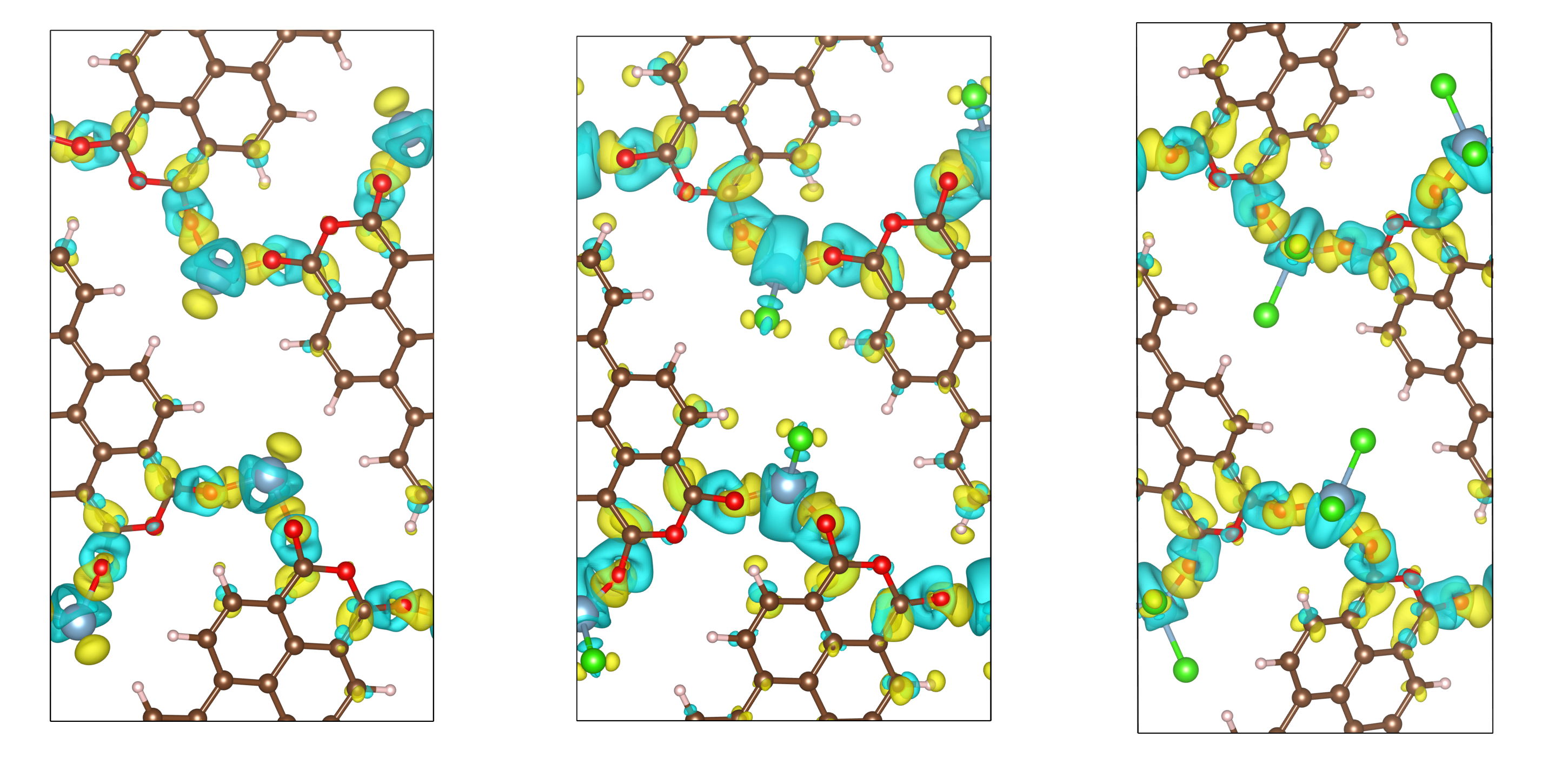}
	\caption{Visualisation of the variations of charge density caused by the insertion of Al$^{3+}$ (left), AlCl$^{2+}$ (centre), AlCl$_2^+$ (right) into $\alpha$-PTCDA using the Bader method, viewed along the \textit{a} axis (colour scheme: C = brown, H = white, O = red, Al = grey, Cl = green). Unit cell sizes are to scale. The yellow surfaces correspond to regions in which charge density was gained (relative to neutrality), while the light-blue surfaces represent regions in which charge density was lost. Isosurface values: 0.005.}
	\label{fig:baders}
\end{figure}

The structural changes resulting from ion intercalation already provide some indication of what the most favourable energy storage mechanism is for the system, as the insertion of AlCl$_x^{(3-x)+}$  ($x$ = 0-2)  ions causes varying degrees of lattice distortion, and the non-covalent intermolecular attractive forces are strongest in the case of AlCl$^{2+}$ intercalation, and and weakest in AlCl$_2^+$- intercalation (Figure S2). Further proof, however, was found by analysing the variation of charge density in the structures using the Bader method.\cite{bader_quantum_1991} In this technique, the charge density distribution of the system is decomposed and assigned to individual atoms on the basis of atomic boundaries, which are constructed by tracing a series of two-dimensional surfaces perpendicular to molecular bonds, on which the charge density is a minimum. Thus, the charge contained in each region represents a good approximation of the total electronic charge of each atom. Finally, by subtracting the charge density of the intercalated structures from that of the non-intercalated ones, we were able to determine which atoms of the structure were most involved in the charge transfer process.
Predictably, it can be seen from Figure \ref{fig:baders} that for all the examined structures structures, the regions in which the vast majority of charge transfer happened are the aluminium atoms and their coordinating carbonyl oxygen atoms. Specifically, the aluminium atoms experience a net loss of charge density, while the oxygen observe a net gain (Table \ref{tab:baders}). However, although it would have been expected for the trivalent Al$^{3+}$ to cause the largest variation of charge density in the structure, this is instead observed in the case of AlCl$^{2+}$ intercalation: each Al atom loses 1.52 electrons, and carbonyl O atoms gain on average 0.52 electrons, compared to -1.09 and +0.37 electrons, respectively, for PTCDA-Al$_2$. Furthermore, in the case of PTCDA-(AlCl)$_2$, the charge is further delocalised onto the perylene backbones, as demonstrated by the presence of yellow regions on most carbon atoms in Figure \ref{fig:baders} (centre). This indicates that, contrary to the other two structures, the aromatic cores of the PTCDA molecules are more prominently involved in the redox mechanism, which constitutes an overall increase in the stability of the system, thanks to a better distribution of electrostatic charges. 
On the other hand, the inverse effect is seen for PTCDA-Al$_2$ (Figure \ref{fig:baders} (left)): the carbon atoms display minimal charge gain, and a sizeable portion of charge density is gained and localised at the tip of the O-Al-O triangles.

\begin{table}[h!]
\caption{Average atomic charge density variation of intercalated PTCDA structures.}
\centering
\begin{tabular}{lcllcllcl}
\toprule
\multicolumn{1}{c}{\textbf{Atom}} & \multicolumn{8}{c}{\textbf{Charge density difference}}                                                           \\ 
	\cmidrule(r){1-9}

                         & \multicolumn{3}{c}{PTCDA-Al$_2$} & \multicolumn{3}{c}{PTCDA-(AlCl)$_2$} & \multicolumn{2}{c}{PTCDA-(AlCl$_2$)$_2$} \\ 
\cmidrule(r){2-9}                         
Al                       & \multicolumn{3}{c}{-1.09}         & \multicolumn{3}{c}{-1.51}             & \multicolumn{2}{c}{-0.91}              \\
O (carbonyl)             & \multicolumn{3}{c}{+0.37}         & \multicolumn{3}{c}{+0.52}             & \multicolumn{2}{c}{+0.35}              \\
C (anhydride)            & \multicolumn{3}{c}{+0.06}         & \multicolumn{3}{c}{+0.13}             & \multicolumn{2}{c}{+0.07}              \\
Cl                       & \multicolumn{3}{c}{-}         & \multicolumn{3}{c}{-0.08}            & \multicolumn{2}{c}{+0.02}     \\       
\bottomrule
\end{tabular}
\label{tab:baders}
\end{table}

Perhaps the most compelling evidence for the determination of the redox mechanism is given by the analysis of the thermodynamics and predicted reaction voltage of each redox process. These were calculated by using the Nernst equation $\Delta _r G^{\standardstate} = - nFE^{\standardstate}$, where  $\Delta  _r G^{\standardstate}$ is the Gibbs free energy of reaction for a given redox process, $n$ is the number of electrons transferred in the reaction, $F$ is the Faraday constant, and $E^{\standardstate}$ is the equilibrium voltage at standard conditions. Because the system is at constant pressure and room temperature, $\Delta  _r G^{\standardstate}$ can be approximated to the energy of reaction $\Delta E_r$, which is obtained by subtracting the total energy of the products to the total energy of the reagents. \cite{he_density_2019,van_der_ven_understanding_2013} Therefore, in order to calculate $\Delta E_r$ for all the investigated processes, the total energy of all the species participating in the redox reactions (Al, AlCl$_4^-$, Al$_2$Cl$_7^-$) was also calculated, using the same parameters used for the intercalated PTCDA structures.  
The results of reaction energy and voltage calculations are found in Table \ref{tab:voltages}. It can be immediately seen that the formation of PTCDA-(AlCl$_2$)$_2$ is not thermodynamically favoured, as it involves a positive $\Delta E_r$ and negative equilibrium voltage. This is consistent with the observations made from examining the structural changes and weakening of non-covalent $\pi$-$\pi$-stacking interactions. Furthermore, while both the formation of PTCDA-Al$_2$ and PTCDA-(AlCl)$_2$ are thermodynamically favoured, the intercalation of AlCl$^{2+}$ shows an equilibrium voltage approximately 200 mV higher than that of Al$^{3+}$. This is a strong indication that AlCl$^{2+}$ is the interacting ion in the redox mechanism, which is in contradiction some of the previous reports, \cite{fang_sulfur-linked_2021} but consistent with others featuring analogue organic cathodes with carbonyl coordinating groups. \cite{bitenc_concept_2020,yoo_tetradiketone_2021}. Even more corroborating evidence was found by calculating the equilibrium voltage of partial intercalation reactions, leading to the formation of PTCDA-(AlCl) and PTCDA-Al: the partial insertion of AlCl$^{2+}$ results in an equilibrium voltage of 0.49 V, a value much higher than the ones obtained from any other calculation. It is worth noting, however, that even this voltage is a relatively poor match with the experimental voltage, which is observed at about 1.05 V. This is likely due to the choice of DFT parameters, which were decided on the basis of a balance between computational time and simulation accuracy. One factor particularly likely to impact the accuracy equilibrium voltage calculations is the energy of the chloroaluminate ions Al$_2$Cl$_7^-$ and AlCl$_4^-$, as the plane-wave basis system used for this work does not accurately represent the energy level of the ions in the solution state. Nevertheless, the large difference in the calculated equilibrium voltages, consistent with the observations made previously, paints a clear picture of what is the most favoured redox mechanism for this system.

\begin {table}[h!]
\caption{Predicted equilibrium voltages for the intercalation of AlCl$_x^{(3-x)+}$ ($x$ = 0-2) ions into PTCDA.}
\centering
\begin{tabular}{lll}
\toprule
\textbf{Reaction}                                                                                         & \multicolumn{1}{l}{$\Delta E_r$ (kJ mol$^{-1}$)} & \multicolumn{1}{l}{\textbf{$E^{\standardstate}$ (V)}} \\ 
	\cmidrule(r){1-3}
PTCDA + 2 Al  $\rightarrow$ \textbf{PTCDA-Al$_2$}                                              & \multicolumn{1}{c}{-43.19}                         & \multicolumn{1}{c}{0.07}                        \\
3 PTCDA + 2 Al$_2$Cl$_7^-$ + 4 Al $\rightarrow$ 3 \textbf{PTCDA-(AlCl)$_2$} + 2 AlCl$_4^-$    & \multicolumn{1}{c}{-303.56}                        & \multicolumn{1}{c}{0.26}                        \\
3 PTCDA + 4 Al$_2$Cl$_7^-$ + 2 Al $\rightarrow$ 3 \textbf{PTCDA-(AlCl$_2$)$_2$} + 4 AlCl$_4^-$ & \multicolumn{1}{c}{91.36}                          & \multicolumn{1}{c}{-0.16}                       \\
PTCDA + Al  $\rightarrow$ \textbf{PTCDA-Al}                                                  & \multicolumn{1}{c}{-15.50}                         & \multicolumn{1}{c}{0.05}                        \\
3 PTCDA +  Al$_2$Cl$_7^-$ + 4 Al $\rightarrow$ 3 \textbf{PTCDA-(AlCl)} +  AlCl$_4^-$           & \multicolumn{1}{c}{-283.02}                       & \multicolumn{1}{c}{0.49}                        \\
\bottomrule
\end{tabular}
\label{tab:voltages}
\end{table}

\begin{figure}[h!]
	\centering
	\includegraphics[width=0.99\textwidth]{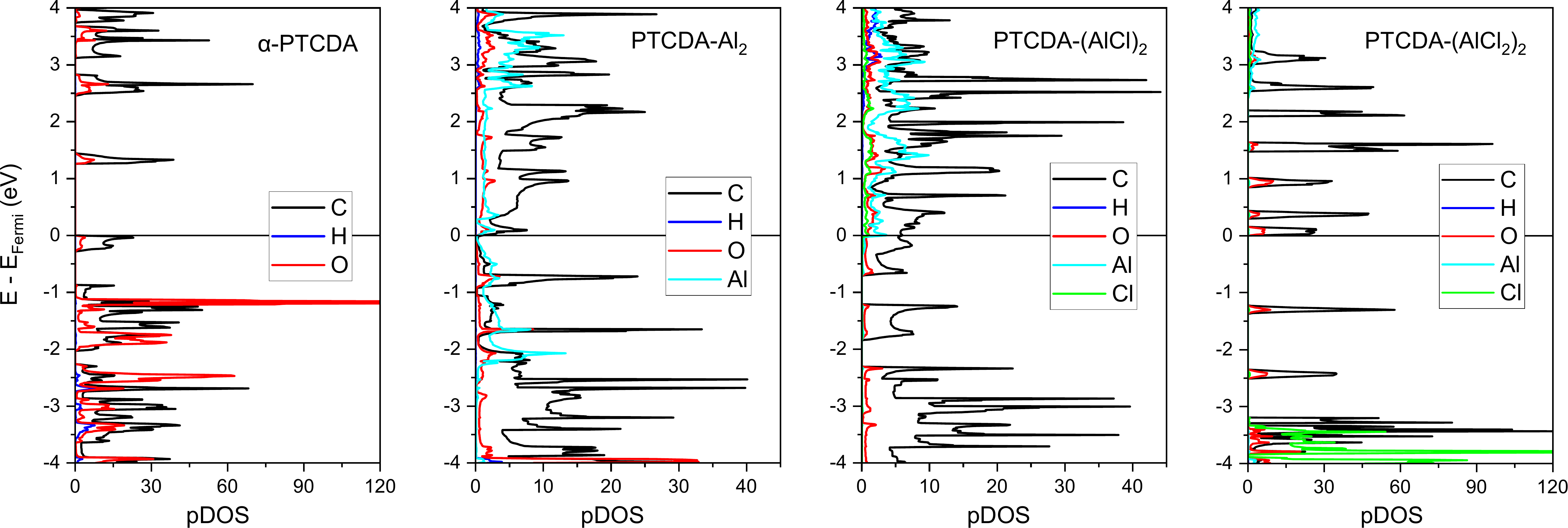}
	\caption{projected densities of states (pDOS) near the Fermi level for $\alpha$-PTCDA and  AlCl$_x^{(3-x)+}$  ($x$ = 0-2) -intercalated structures.}
	\label{fig:pdos}
\end{figure}

In order to further characterise and evaluate the properties of the systems, the projected densities of states (pDOS) and band diagrams for the starting $\alpha$-PTCDA structure and the AlCl$_x^{(3-x)+}$  ($x$ = 0-2) -intercalated structures were calculated (Figures S3-S8). It can be seen from Figure \ref{fig:pdos} that, predictably, the calculated pDOS for $\alpha$-PTCDA presents a series of isolated bands, with a gap of about 1.2 eV between the highest occupied and lowest unoccupied bands. This is consistent with the molecular character of PTCDA and its known behaviour as an organic semiconductor. Conversely, both PTCDA-Al$_2$ and PTCDA-(AlCl)$_2$ present a gap-less or near-gap-less band structure, similar to what was reported in previous publications for analogue systems.\cite{zhou_high-capacity_2021} The AlCl-intercalated system, in particular, presents a broad, continuous non-zero DOS starting from about -0.7 eV vs. the Fermi level; these states are predominantly localised on the carbon atoms, which is consistent to the observations made from the Bader charge analysis. While the complete lack of a band gap may not be an accurate representation of the energy levels of the system, due to the known tendency of DFT to underestimate band gaps, this result indicates that AlCl-intercalated PTCDA has an enhanced electronic conductivity, which is beneficial to its performance as a battery cathode. Finally, the band structure of PTCDA-(AlCl$_2$)$_2$ presents again a series of isolated states with a highest occupied-lowest unoccupied gap of approximately 1.2 eV; this is consistent with previous observations made about this system, making it the least likely among the examined redox processes to be taking place.

\section{Confirming with experimental data}
\begin{figure}[h]
	\centering
	\includegraphics[width=0.99\textwidth]{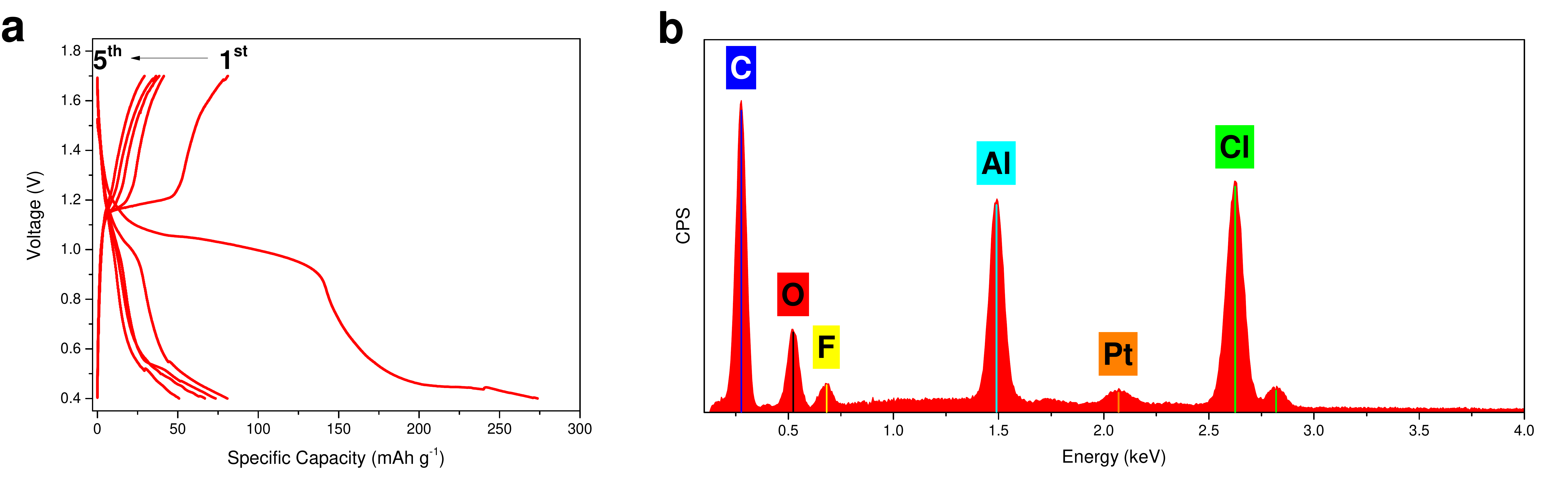}
	\caption{(a) Galvanostatic profiles of the first 5 discharge-charge cycles of an aluminium-ion battery built using PTCDA as cathode. (b) Energy disperse X-ray (EDX) spectrum of the PTCDA cathode discharged to 0.4 V
	. the F and Pt peaks are attributed to the fluorinated binder polymer and conductive coating required for SEM imaging, respectively.}
	\label{fig:exps}
\end{figure}

Galvanostatic discharge/charge tests were performed on prototype battery devices using PTCDA as cathode. It can be seen from Figure \ref{fig:exps}a that the discharge profile of PTCDA presents a first plateau around 1.05 V, terminating in correspondence to a specific capacity of approximately 144 mAh g$^{-1}$. Based on the relationship between theoretical capacity and electron transfer $C = \frac{nF}{3.6m}  $, where $C$ is the specific capacity, $F$ is the Faraday constant, and $m$ is the molecular mass of PTCDA, we can infer that discharging the cathode to 0.8 V corresponds to a transfer of about two electrons per formula unit. This result is therefore a good indication that this plateau corresponds to the intercalation of AlCl$^{2+}$ to form PTCDA-(AlCl), for which the theoretical specific capacity is 136.64 mAh g$^{-1}$. A second discharge plateau at about 0.45 V can also be seen in Figure \ref{fig:exps}a. This likely corresponds to the intercalation of the second AlCl$^{2+}$ ion per formula unit, leading to the formation of PTCDA-(AlCl)$_2$. While the measured specific capacity does indeed match well with the predicted theoretical value of 273.27 mAh g$^{-1}$, this reaction appears to be irreversible due to the absence of a corresponding plateau in the charging step; furthermore, this plateau could also be imputable to secondary contributions to discharge capacity caused by the metallic current collector. \cite{shi_avoiding_2019} It is also worth noting that the coulombic efficiency of the discharge/charge cycle does not reach satisfactory values for a commercially viable device: this is a known issue due to solubility of PTCDA in the chloroaluminate electrolyte, which could be improved by chemical modifications of the cathode such as polymerisation or functionalisation.

Energy disperse X-ray (EDX) spectroscopy was performed on the discharged cathode to assess its elemental composition (Figure \ref{fig:exps}). It can be that both Al and Cl are clearly present in the sample, which provides further evidence of either AlCl$^{2+}$ or AlCl$_2^+$ being the intercalating species. Looking at atomic percentages, the two elements are present in a 1:1.58  Al:Cl ratio. While this does not match well with either intercalating species, many factors could be responsible for this discrepancy; the most likely cause, however, is the presence of leftover chloroaluminate ions (AlCl$_4^-$, Al$_2$Cl$_7^-$) in the sample, resulting in a higher Al:Cl ratio than expected by the presence of intercalated ions exclusively. Therefore, the EDX spectrum is further proof that the redox mechanism involves the intercalation of AlCl$^{2+}$ ions. Finally, additional evidence disproving the intercalation of Al$^{3+}$ can be found in the comparison between the predicted infrared spectrum for PTCDA-Al$_2$ and the measured spectrum for the discharged cathode (Figure S9). 


In conclusion, we have investigated the redox mechanism for non-aquoeus Al-ion batteries employing PTCDA as cathode. Using a combination of DFT and experimental data, we have demonstrated that the energy storage mechanism involves the reversible transfer of two electrons per formula unit, and a likely irreversible transfer of two further electrons at lower voltage, thanks to the intercalaton of AlCl$^{2+}$ ions, which are complexed by two carbonyl groups belonging to adjacent PTCDA molecules. With this work, we have therefore disproved previous claims that this molecule allows the intercalation of trivalent aluminium ions, thus aligning the behaviour of the cathode with the trend observed for other organic compounds employing carbonyl groups as the main redox-active site.
While there are still some significant challenges in the use of PTCDA for commercialisation-ready devices, mainly due to its high solubility and sub-optimal energy density, this paper hopefully provides some clarity over the mechanism enabling organic carbonyl compounds to function as cathodes for chloroaluminate aluminium-ion batteries. We hope that this piece of knowledge will help contributing in the creation of high-energy density electrode materials in the future. 

\section{Materials and methods}
\label{experimental}
\subsection{Computational details}
DFT data was obtained using the Quantum ESPRESSO package.\cite{giannozzi_quantum_2009,giannozzi_advanced_2017}. Calculations were executed using Perdew, Burke and Ernzerhof's formulation of the generalised gradient approximation (GGA) \cite{perdew_generalized_1996} of exchange-correlation energy of electrons. Projector augmented-wave (PAW) \cite{blochl_projector_1994} pseudopotentials were employed to describe interactions between core and valence electrons, using an energy cut-off of 60 Ry for the wavefunction, and 600 Ry for the electron density. All structures were relaxed to a total force lower than 10$^{-4}$ Ry/Bohr. A uniform 10 $\times$ 3 $\times$ 2 Monkhorst-Pack k-point mesh was used to sample the Brillouin zone for relaxation and self-consistent energy field calculations, while a finer 30 $\times$ 9 $\times$ 6 grid was used for DOS and pDOS calculations. Grimme's DFT-D3 \cite{grimme_consistent_2010} approximation was implemented to account for the significant long range non-covalent interactions present in the system. For prediction of FTIR spectra, structures were re-optimised using Grimme's DFT-D correction for non-covalent interactions,\cite{grimme_semiempirical_2006} and using fixed electronic occupations. For calculations of chloroaluminate ions, only the $\Gamma$ k-point was sampled; the ions were placed in a cubic cell with a lattice parameter of 50 Bohr to avoid intermolecular interaction, and the Martyna-Tuckerman correction to total energy and self consistent field potential for isolated systems \cite{martyna_reciprocal_1999} was applied. The Bader charge analysis was performed using the code developed by the Henkelman group.\cite{tang_grid-based_2009,yu_accurate_2011} In order to identify high-symmetry k-paths in the Brillouin zone for the calculation of band structures, the online tool SeeK-Path was used,\cite{hinuma_band_2017} employing the \texttt{spglib} software library \cite{togo_textttspglib_2018} to identify crystallographic symmetry. Visualisations were created using the VESTA software.

\subsection{Preparation of PTCDA cathodes}
3,4,9,10-Perylenetetracarboxylic dianhydride (PTCDA) (Sigma-Aldrich) was annealed at 450 \degree  C for 4 hours under nitrogen atmosphere to promote partial oligomerisation of molecular units and prevalent formation of the $\alpha$-polymorph.\cite{fan_ultrafast_2018,huang_gas_2012} Then, a slurry was formed by mixing the annealed PTCDA with carbon black (Super-P conductive, Alfa Aesar) and PVDF as binder (average M$_w$ 534,000, Sigma-Aldrich) in a 8:1:1 weight ratio in 1-methyl-2-pyrrolidone (Sigma-Aldrich). The slurry was coated on a Molybdenum current collector using a blade coater, and dried at 80 \degree C in a vacuum oven. after drying, the coated foils were cut into 11 mm-diameter disks and used as cathodes in the construction of battery devices.

\subsection{Preparation of electrolytes}
1-ethyl-3-methylimidazolium (EMIm) chloride (Sigma-Aldrich) and Aluminium trichloride (Sigma-Aldrich) were gradually mixed in a 1:1.3 molecular ratio under magnetic stirring in a nitrogen-filled glovebox. An exothermic reaction takes place, leading to the formation of a pale yellow liquid. the mixture was left stirring for at least two hours before being used in the battery devices.

\subsection{Assembly of battery devices}
Custom-built Swagelok-type cells consisting of a cylindrical PEEK body with an inner diameter of 12 mm and two Molybdenum rods as current collector were used as casings for the battery devices. All components were initially baked at 100 \degree C under vacuum for at least 2 hours to remove any residual water, then immediately transferred into a nitrogen-filled glovebox with the oxygen and water levels kept below 1 ppm. PTCDA-coated Mo foils were used as cathodes, and 11 mm diameter, high-purity Al foil discs (11 mm diameter) were used as anodes. Glass microfibre (GF/D) discs with 12 mm diameters, soaked in approximately 120 µL of the electrolyte, were used as separators. The devices were then wrapped with Parafilm as an additional moisture barrier, and taken outside the glovebox for electrochemical testing.

\subsection{Characterisation}
X-ray diffractogram were acquired with a PANalytical Xpert XRD device, using a Cu K$\alpha$ source. SEM images and EDX spectra were acquired using a ZEISS Sigma VP SEM using an accelerating voltage of 15 kV. Samples were sputter coated with Pt to allow electric conductivity and enhance image quality. for \textit{ex-situ} analysis, discharged battery devices were taken disassembled inside the glovebox, the cathodes were rinsed with dichloromethane, and kept inside the glovebox until immediately before the measurement to limit ambient exposure. Galvanostatic charge-discharge  experiments were performed using a NEWARE BTS CT-4008-5V10mA-164 battery analyser system. FTIR spectra were acquired using a PerkinElmer Spectrum II spectrophotometer equipped with a UATR accessory.



\bibliographystyle{ieeetr}
\bibliography{references}  

\end{document}


\beginsupplement

\title{Unraveling the multivalent Aluminium-ion intercalation mechanism in 3,4,9,10-Perylenetetracarboxylic dianhydride (PTCDA)}
\maketitle

\author{Nicolò Canever and Thomas Nann}
\date{}

\section*{Supporting information}

\begin{figure}[h!]
	\centering
	\includegraphics[width=0.29\textwidth]{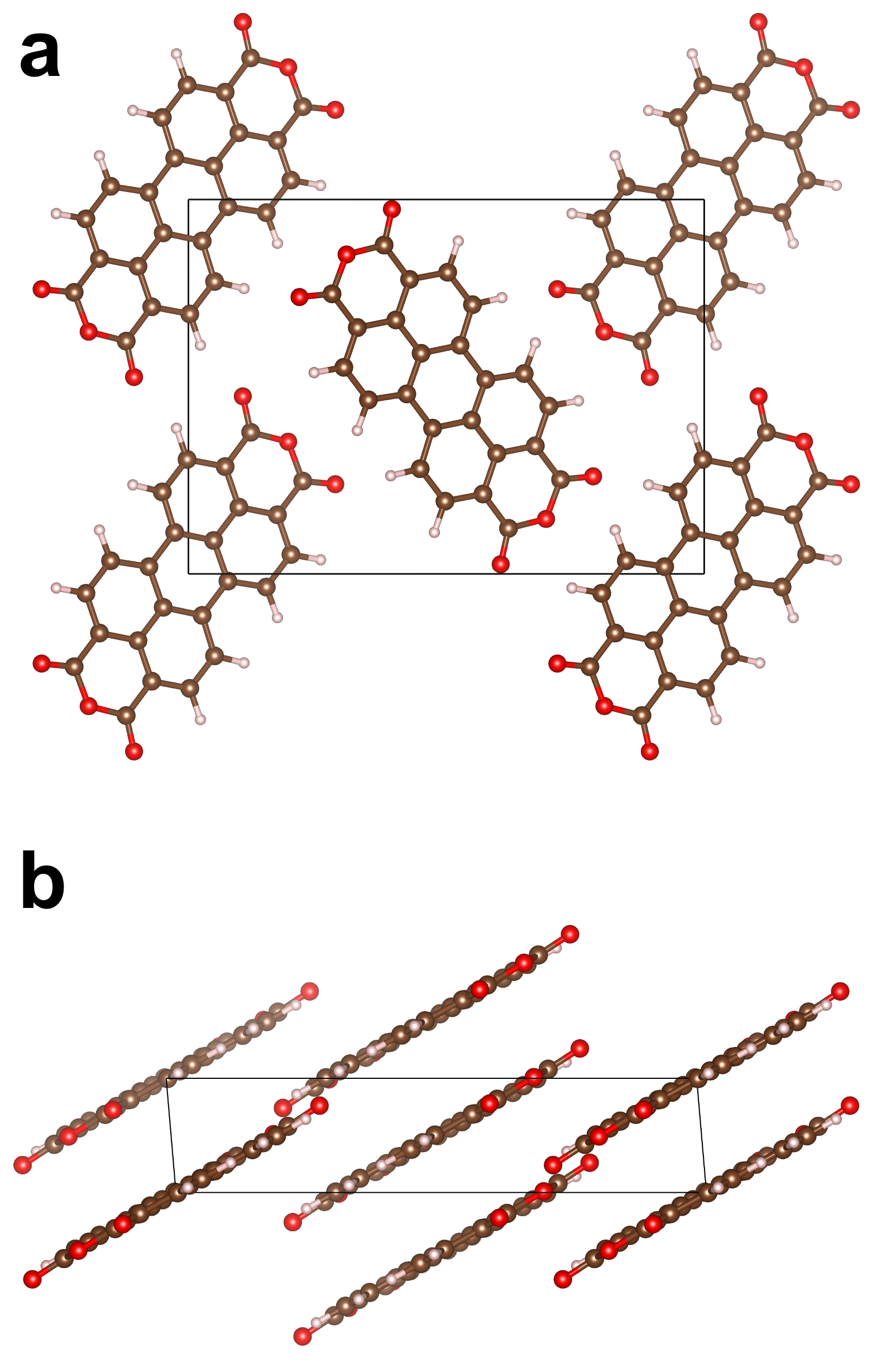}
	\includegraphics[width=0.65\textwidth]{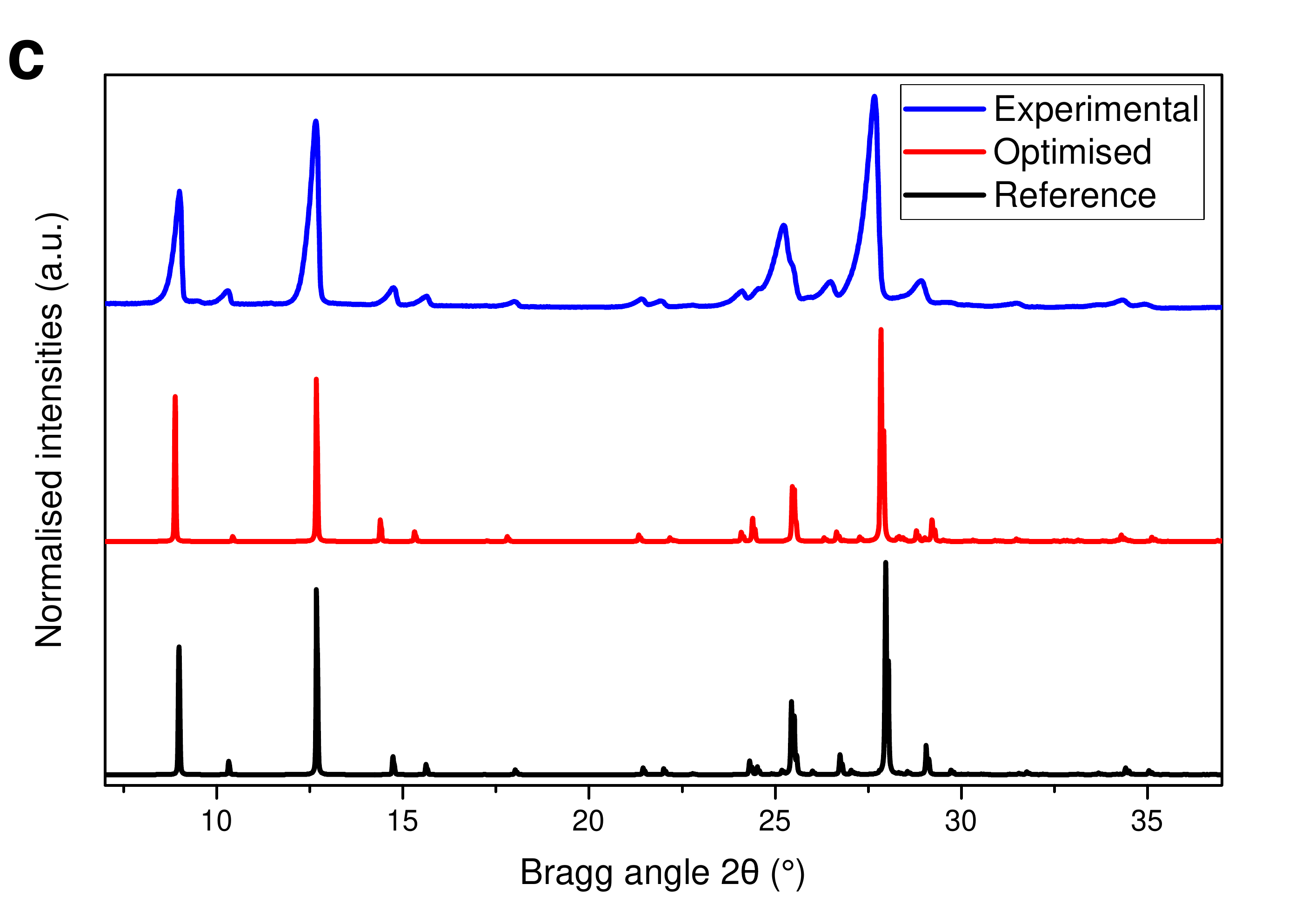}
	\caption{(a,b) Optimised geometry of the $\alpha$-PTCDA unit cell, projected along the \textit{a} (top) and \textit{b} (bottom) crystallographic axes (colour scheme: C = brown, H = white, O = red). The individual PTCDA molecules are arranged in the typical 'herringbone' configuration and the molecular planes are all parallel, which is consistent with the known structure of the $\alpha$-polymorph.  (c) Comparison between the X-ray diffraction data predicted by the literature reference, \cite{tojo_refinement_2002} the relaxed structure calculated by Density Functional Theory (DFT), and the experimental powder diffraction data of the PTCDA sample used in this publication. }
	\label{fig:ptcda-pristine-xrd}
\end{figure}

\begin{figure}[h!]
	\centering
	\includegraphics[width=0.3\textwidth]{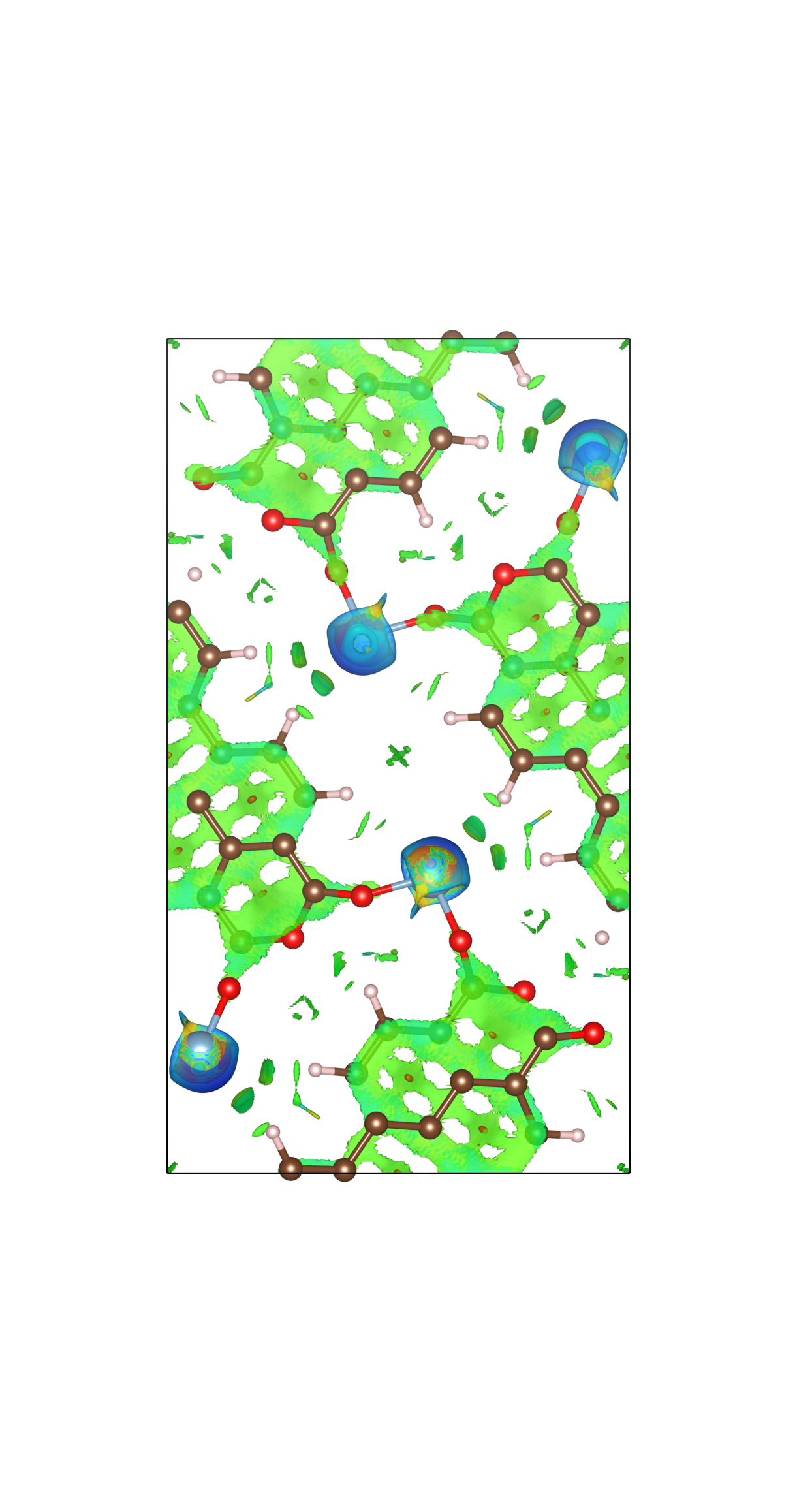}
	\includegraphics[width=0.3\textwidth]{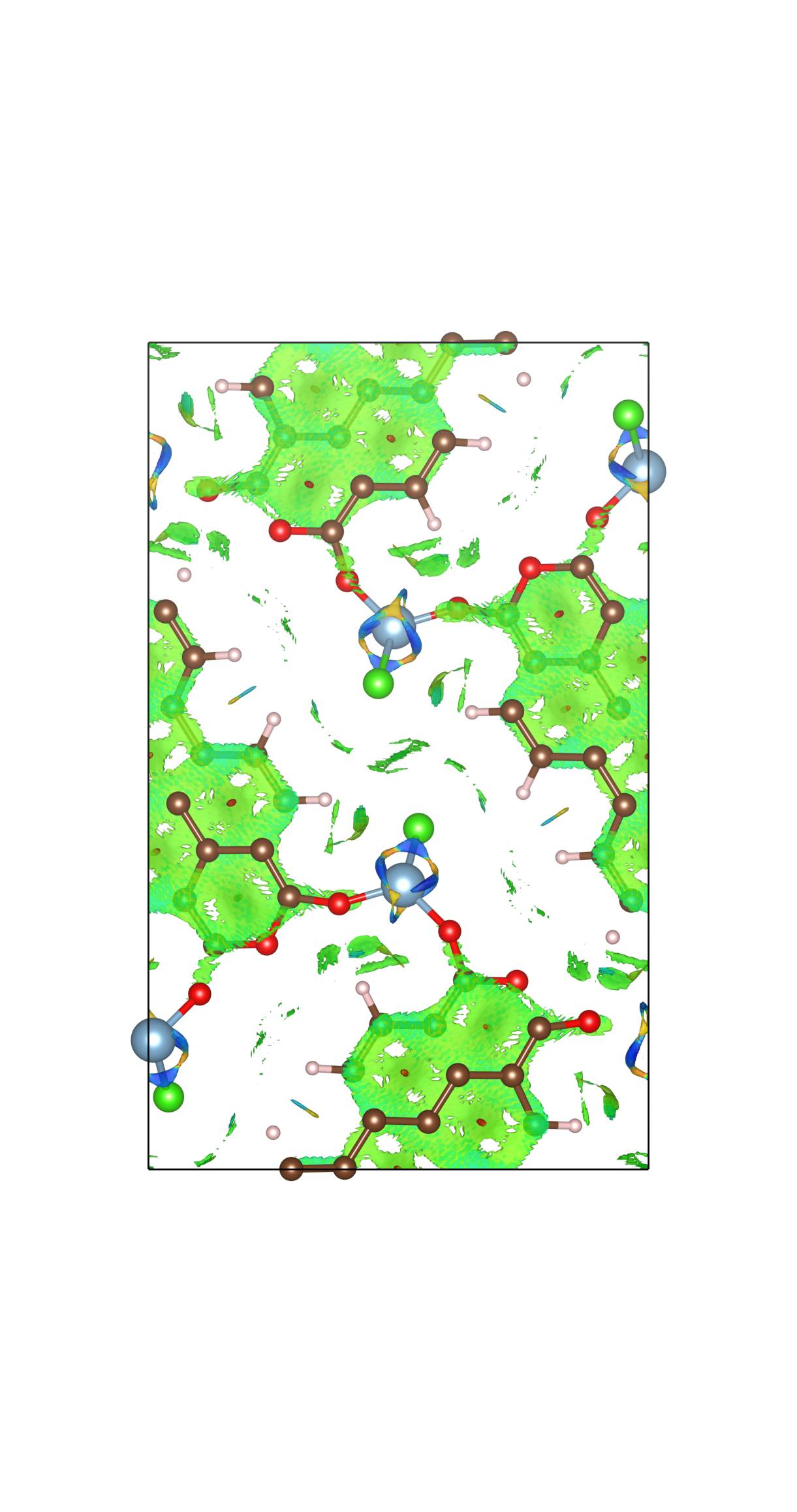}
	\includegraphics[width=0.3\textwidth]{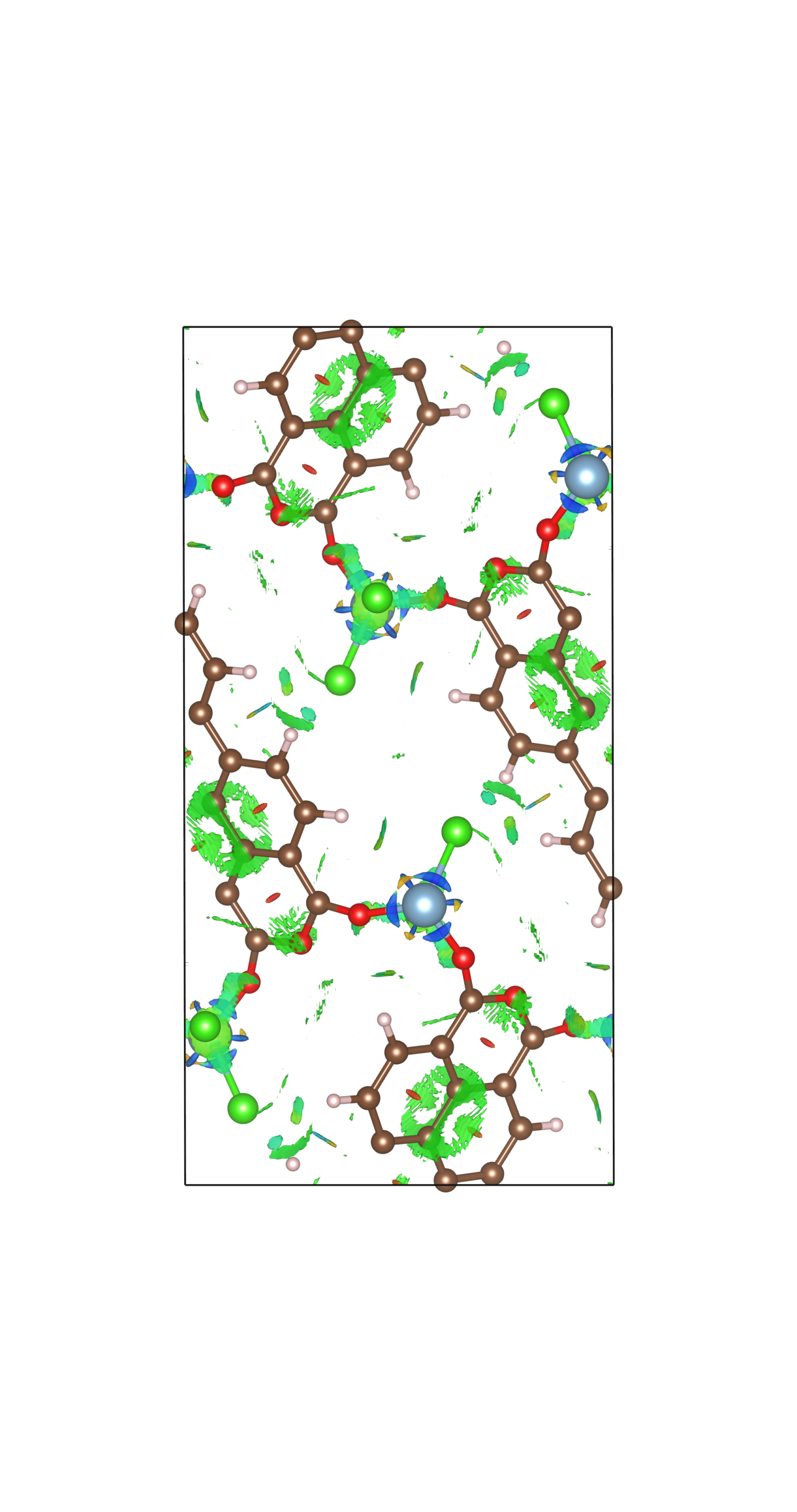}
	\caption{Visualisation of non-covalent interactions of PTCDA intercalated with Al$^{3+}$ (left), AlCl$^{2+}$ (center), AlCl$_2^+$ (right), using the method developed by Contreras-García \textit{et al}.\cite{contreras-garcia_nciplot_2011} The green regions overlapping the aromatic regions of the PTCDA molecules represent $\pi$-$\pi$-stacking interactions in the lattice. It can be seen that the AlCl$^{2+}$-intercalated structure presents the strongest interactions, as indicated by the fuller regions. On the other hand, the intermolecular interactions are greatly reduced in the case of AlCl$_2^+$ intercalation.   Isosurfaces value: 0.3.}
	\label{fig:noncovalent}
\end{figure}

\begin{figure}[h!]
	\centering
	\includegraphics[width=0.99\textwidth]{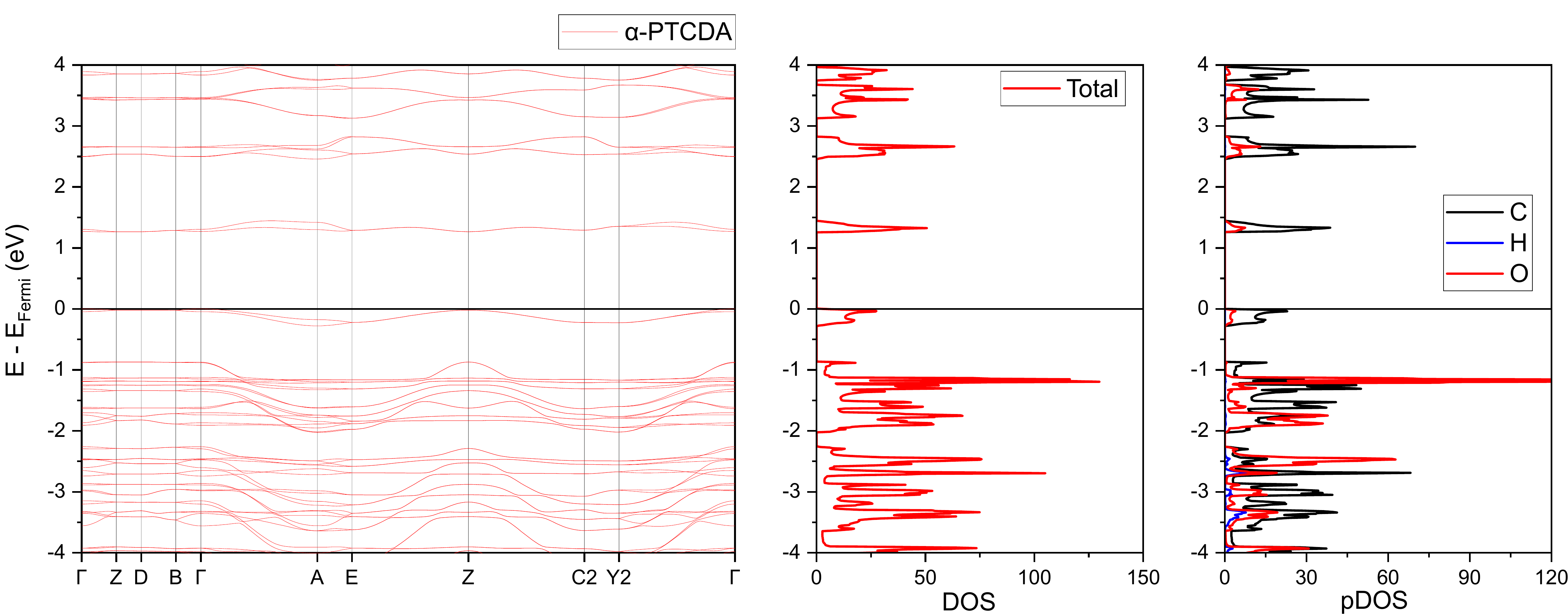}
	\caption{Band diagram, density of states (DOS) and projected density of states (pDOS) for the calculated structure of $\alpha$-PTCDA.}
	\label{fig:noncovalent}
\end{figure}

\begin{figure}[h!]
	\centering
	\includegraphics[width=0.99\textwidth]{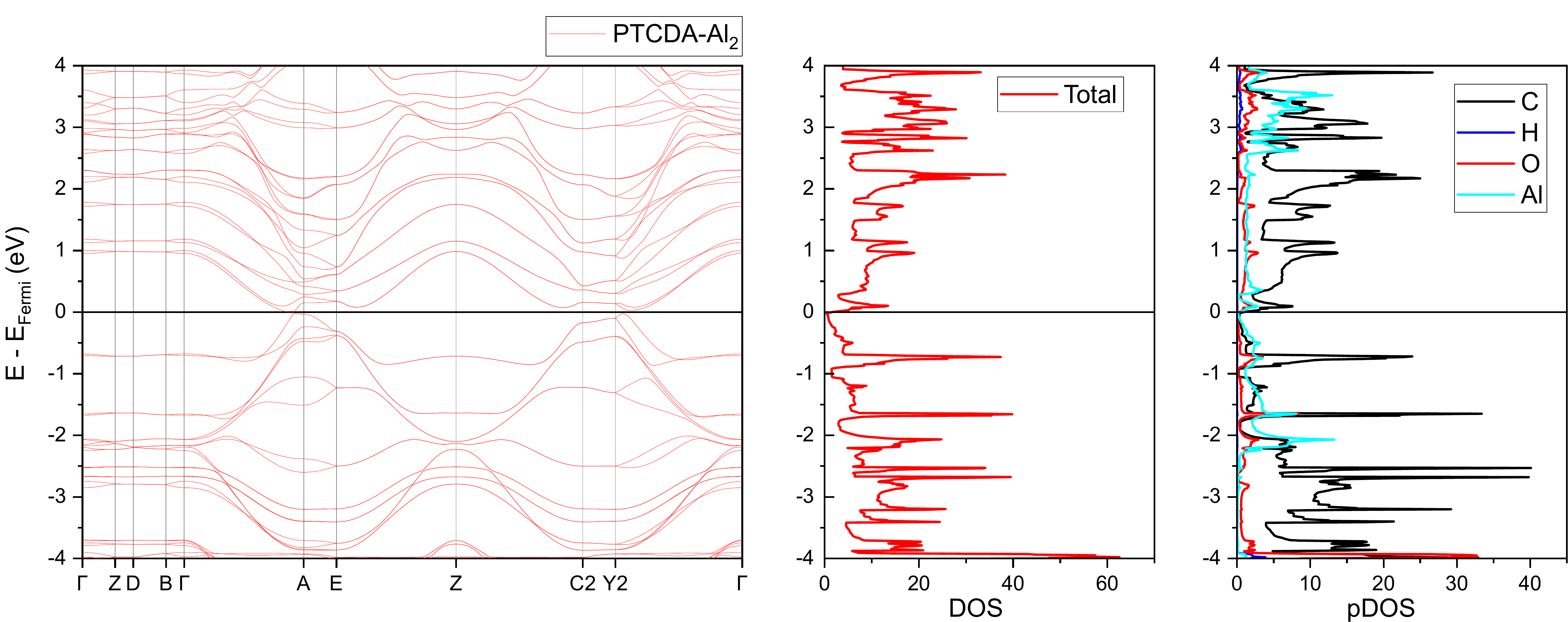}
	\caption{Band diagram, density of states (DOS) and projected density of states (pDOS) for the calculated structure of PTCDA-Al$_2$.}
	\label{fig:al-bands}
\end{figure}

\begin{figure}[h!]
	\centering
	\includegraphics[width=0.99\textwidth]{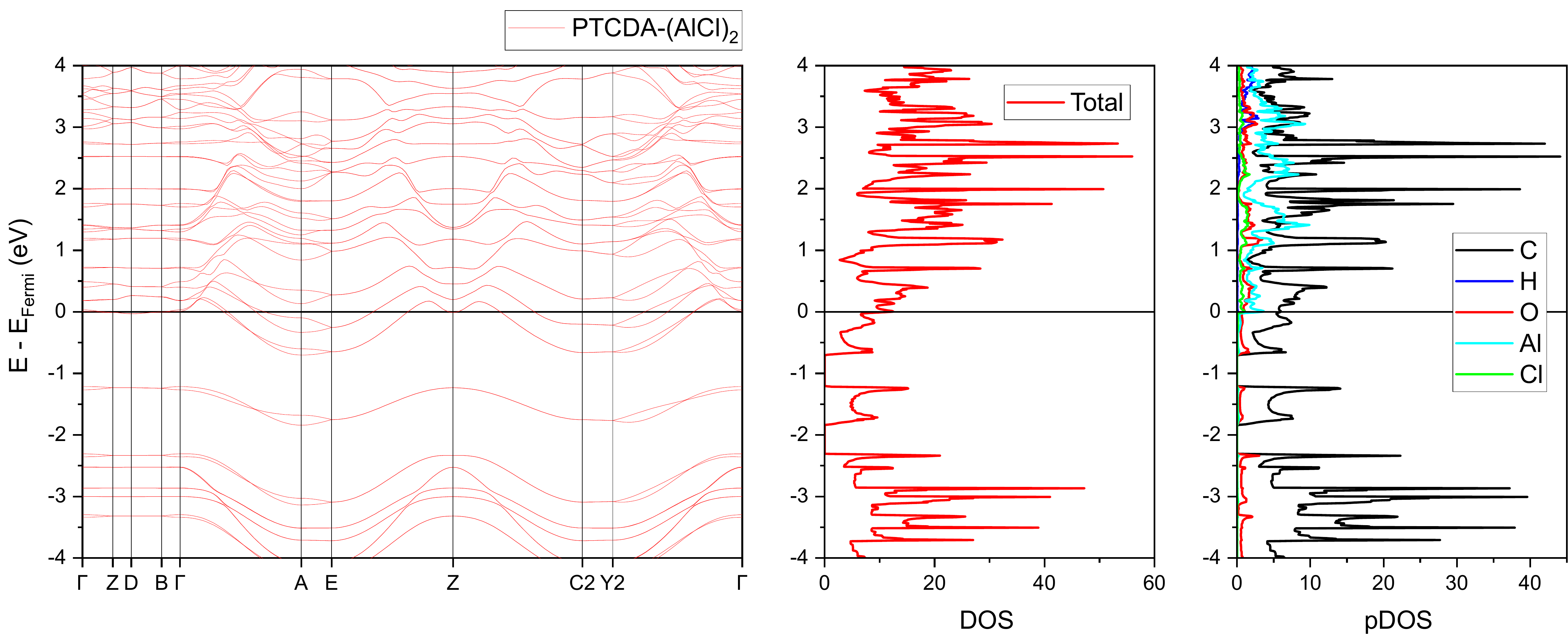}
	\caption{Band diagram, density of states (DOS) and projected density of states (pDOS) for the calculated structure of PTCDA-(AlCl)$_2$.}
	\label{fig:alcl-bands}
\end{figure}

\begin{figure}[h!]
	\centering
	\includegraphics[width=0.99\textwidth]{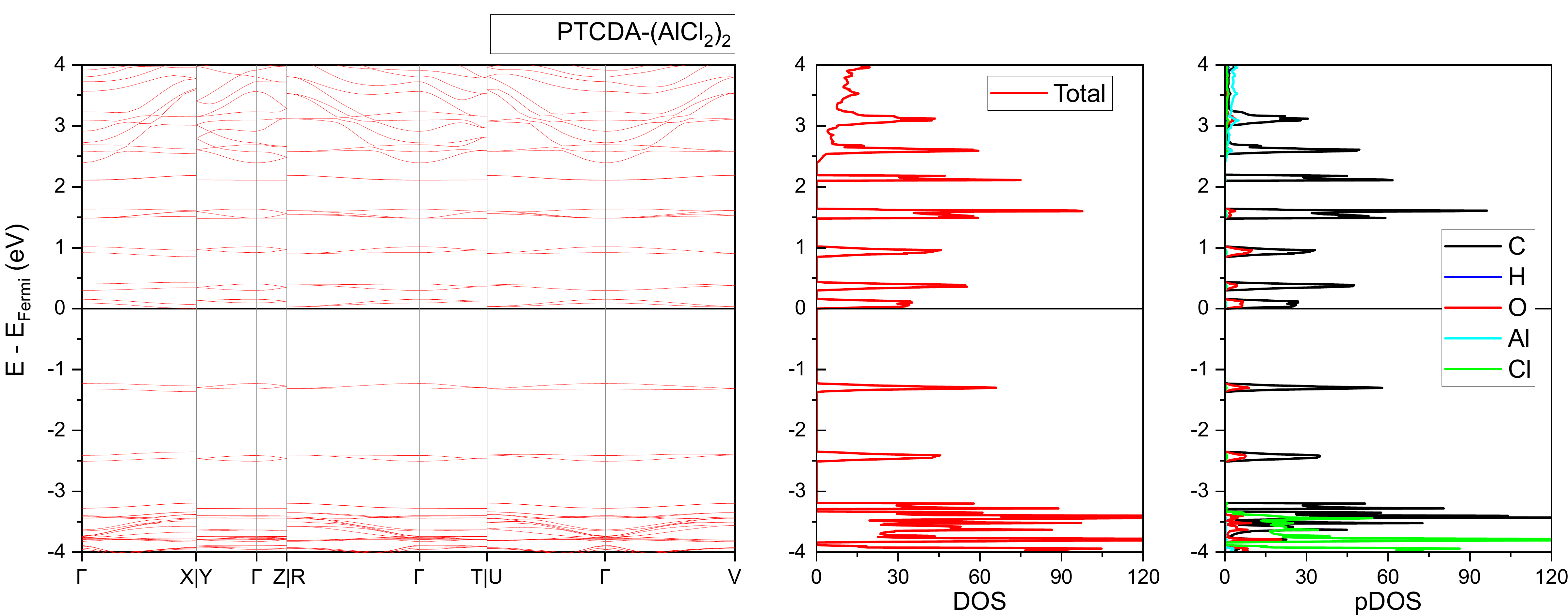}
	\caption{Band diagram, density of states (DOS) and projected density of states (pDOS) for the calculated structure of PTCDA-(AlCl$_2$)$_2$.}
	\label{fig:alcl2-bands}
\end{figure}

\begin{figure}[h!]
	\centering
	\includegraphics[width=0.99\textwidth]{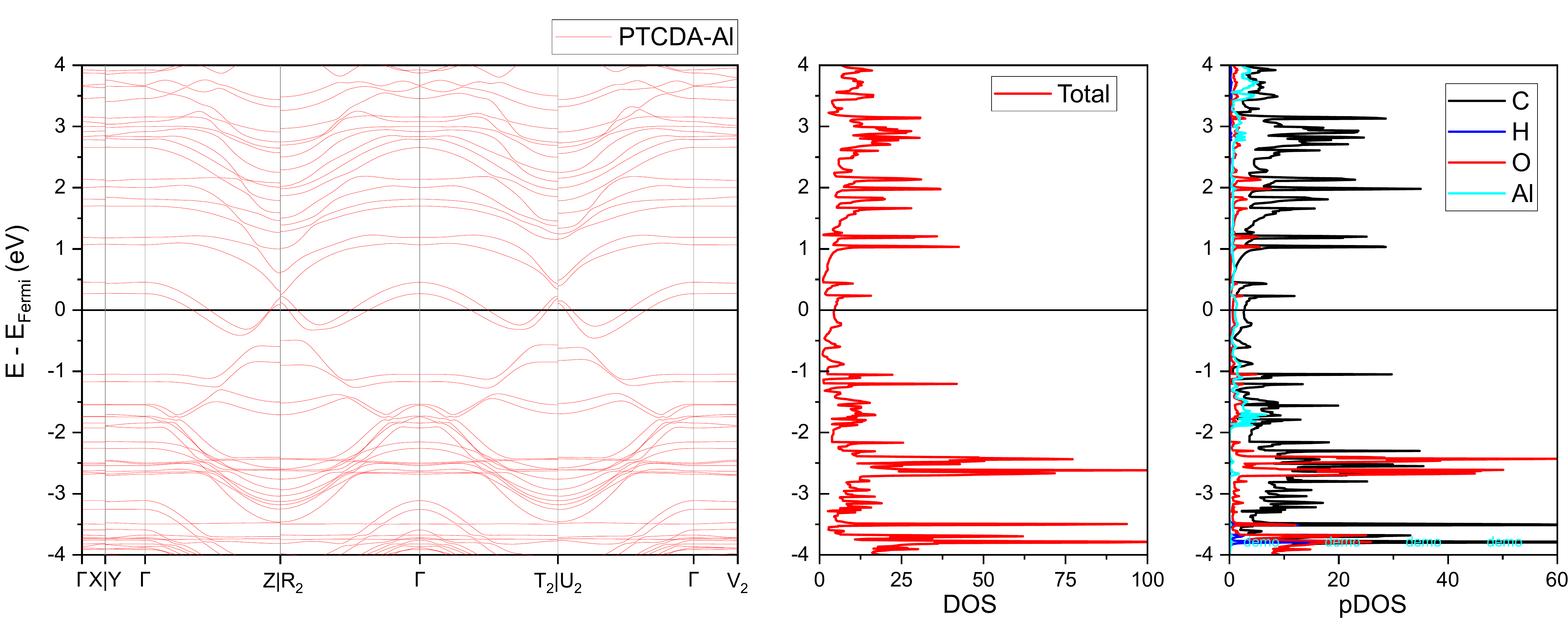}
	\caption{Band diagram, density of states (DOS) and projected density of states (pDOS) for the calculated structure of the partially intercalated structure PTCDA-Al.}
	\label{fig:al-half-bands}
\end{figure}

\begin{figure}[h!]
	\centering
	\includegraphics[width=0.99\textwidth]{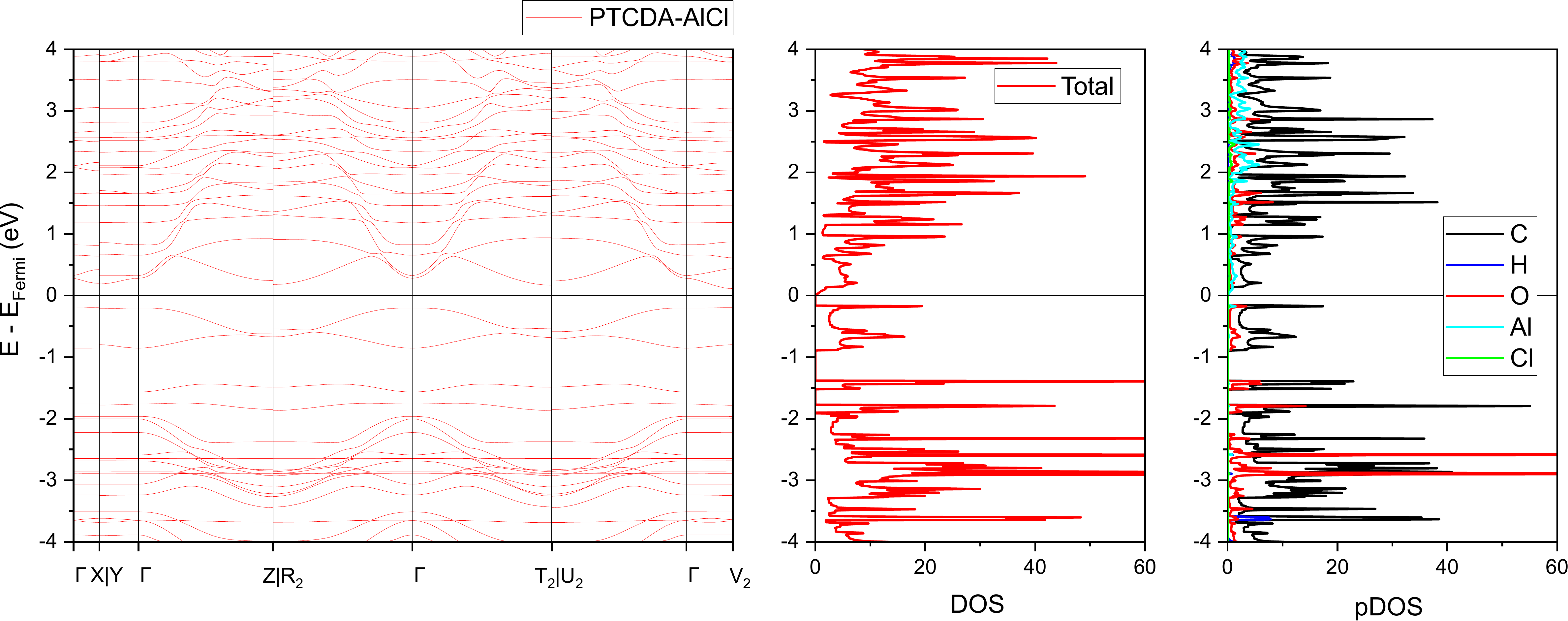}
	\caption{Band diagram, density of states (DOS) and projected density of states (pDOS) for the calculated structure of the partially intercalated structure PTCDA-AlCl.}
	\label{fig:alcl-half-bands}
\end{figure}

\begin{figure}[h!]
	\centering
	\includegraphics[width=0.7\textwidth]{ptcda ftir.pdf}
	\caption{Band diagram, density of states (DOS) and projected density of states (pDOS) for the calculated structure of the partially intercalated structure PTCDA-AlCl.}
	\label{fig:ftir}
\end{figure}

\clearpage
\bibliographystyle{ieeetr}
\bibliography{references}